1**Synthesis of nanocrystalline δ-MoN by thermal annealing of amorphous thin films grown on (100) Si by reactive sputtering at room temperature**

N. Haberkorn,[1,2] S. Bengio,[1] H. Troiani,[1,2] S. Suárez,[1,2] P. D. Pérez,[1] M. Sirena,[1,2] J. Guimpel.[1,2]

[1]*Centro Atómico Bariloche, Comision Nacional de Energía Atómica, Consejo Nacional de Investigaciones Científicas y Técnicas, Av. Bustillo 9500, 8400 San Carlos de Bariloche, Argentina.*

[2] *Instituto Balseiro, Universidad Nacional de Cuyo, Comision Nacional de Energía Atómica, Av. Bustillo 9500, 8400 San Carlos de Bariloche, Argentina*We report on the synthesis and characterization of nanocrystalline δ-MoN by crystallization of amorphous thin films grown on (100) Si by reactive sputtering at room temperature. Films with chemical composition MoN were grown using a deposition pressure of 5mTorr with a reactive mixture of Ar/(Ar+$N_2$)=0.5. The as-grown films display mostly amorphous structure. Nanocrystalline δ-MoN phase is obtained after annealing at temperatures above 600 °C. The superconducting critical temperature $T_c$ depends on film thickness. Thick films (170 nm) annealed at 700 °C for 30 min display a $T_c$ = 11.2 K (close to the one reported for bulk specimens: 13 K), which is gradually suppressed to 7.2 K for 40 nm thick δ-MoN films. Our results provide a simple method to synthesize superconducting nitride thin films on silicon wafers with $T_c$ above the ones observed for conventional superconductors such as Nb.

## 1. *Introduction*

The synthesis of superconducting transition-metal nitrides is of technological relevance in the design of devices such as tunnel junctions [1] and electromagnetic radiation detectors [2,3]. The molybdenum nitrides present several superconducting crystalline phases: γ-$Mo_2N$ (cubic) with $T_c$ ~ 5 K [4], β-$Mo_2N$ (tetragonal) with superconducting critical temperature $T_c$ ~ 5 K [5] and δ-MoN (hexagonal) with $T_c$ ~ 13 K [6,**7**]. Thin films of the different crystalline structures have been sintered through chemical and physical routes [8,9,10,11,12,13]. An outstanding feature for γ-$Mo_2N$ and δ-MoN is that sharp superconducting transitions have been observed for epitaxial and polycrystalline thin films [14,15,16]. Moreover, superconductivity has been observed for films with thicknesses of a few nanometers [14,15,17]. Recently, we have reported that the chemical concentration and $T_c$ for Mo-N thin films grown at room temperature by reactive sputtering can be modified by changing the Ar:$N_2$ ratio [18]. Films grown with $N_2/(Ar+N_2)$ < 0.4 mixtures are superconducting with nanocrystalline γ-$Mo_2N$ phase. Films grown with $N_2/(Ar+N_2)$ > 0.4 mixtures are mostly amorphous with a stoichiometry Mo/N ≈ 1 (close to the one corresponding to δ-MoN). These amorphous films do not display superconducting transition. It is worth noting that the δ-MoN phase displays $T_c$ higher than Nb (hence, displays larger superconducting gap Δ = 1.76*$k_B$*$T_c$). In addition, hexagonal δ–MoN is the hardest superconductor metal nitride (relevant for coating) [7].

In this work, we report the synthesis of nanocrystalline δ-MoN by crystallization of amorphous thin films. Initially, stoichiometric MoN thin films are grown by reactive DC sputtering at room temperature on AlN buffered (100) Si substrates using a $N_2/(Ar+N_2)$= 0.5 mixture. As-grown thin films display a residual-resistivity ratio $RRR = R^{300K}/R^{onset}$ ≈ 0.8. Nanocrystalline δ-MoN phase with $RRR$ > 1 is obtained after annealing at temperatures above 600 °C. The annealed films display very smooth surfaces. The superconducting critical temperature $T_c$ is strongly affected by the film thickness. The successful synthesis of δ-MoN thin films on Si substrates by a controllable and reproducible route (with low roughness values), enhances the potential applications in superconducting devices.

## 2. *Material and methods*

MoN films were deposited on 8 nm thick AlN buffered Si (100) substrates by reactive DC magnetron sputtering using a $N_2/(Ar+N_2)= 0.5$ mixture without any intentional heating of the substrate [17,18]. The AlN buffer layer was introduced to avoid any chemical reaction between the Mo and the $SiO_2$ cup layer of the Si wafers. AlN was selected because it displays high stability at interfaces with transition metal nitrides [1]. However, no appreciable differences in $T_c$ were observed for thin films grown with and without the buffer layer (not shown). The residual pressure of the chamber was less than $10^{-6}$ Torr. Ultra-high purity Ar (99.999%) and $N_2$ (99.999%) were used as gas sources. The AlN and MoN layers were grown by RF (100 W) and DC (50 W) magnetron sputtering, respectively. During deposition the target to substrate distance was ~ 5.5 cm. The total pressure at the chamber was 5 mTorr. AlN buffer layers were grown using a $N_2/(Ar+N_2)= 0.2$ mixture [17]. The deposition rate for MoN using a $N_2/(Ar+N_2)= 0.5$ mixture was $\approx$ 17 nm / min. Thermal annealing was performed in vacuum (to avoid surface contamination) with a residual pressure of $10^{-5}$ Torr at temperatures between 400 °C and 700 °C during periods of 30 min. In order to homogenize the temperature, the films were enveloped in a tantalum foil during the annealing procedure. Wherever used, the notation [*d*-T] corresponds to samples with thickness *d* annealed at *T* °C (with *T* = 400 °C, 600 °C and 700°C) for 30 min. The results shown here correspond to 40 nm, 60 nm, 100 nm and 170 nm thick MoN films. However, most studies are performed in 170 nm thick MoN films, which display the highest $T_c$.

X-ray diffraction (XRD) patterns were obtained with a Panalytical Empirean equipment. The structure was studied by transmission electron microscopy (TEM) with a Philips CM200UT microscope operating at 200 kV. The TEM specimen was prepared by scraping the surface of a film with a diamond tip. The topology of the films was characterized by atomic force microscopy (AFM) measurements in a Dimension 3100©Brucker microscope. The AFM images were obtained in tapping mode. The chemical composition and thickness of the films were analyzed by Rutherford Backscattering Spectroscopy (RBS) with a

TANDEM accelerator using a 2 MeV $4He^{2+}$ ion beam. Surface composition analysis was performed by X-ray photoelectron spectroscopy (XPS) using a standard Al/Mg twin-anode X-ray gun and a hemispherical electrostatic electron energy analyzer (high vacuum conditions with a base pressure of $10^{-9}$ Torr). The electrical transport measurements were performed using the standard four-point configuration. Magnetization measurements were performed in a SQUID magnetometer. Critical current densities ($J_c$) were estimated considering the Bean model [19].

## 3. Results and discussion

The chemical composition of the films was verified using RBS. The composition $MoN_{(1.00\pm0.05)}$ was observed for pristine and annealed specimens. As-grown MoN thin films using a $N_2/(Ar+N_2) = 0.5$ mixture are mostly amorphous (a little reflection attributed to γ $Mo_2N$ is observed) [18]. The crystallization of δ-MoN after thermal annealing at temperatures above 600 °C was confirmed by XRD (see Fig. 1). The lattice parameters calculated from the (002) and (200) diffraction peaks, are $a = 0.572(2)$ nm and $c = 0.555(3)$ nm. These values are close to those observed in epitaxial δ-MoN [12]. Figures 2*abc* show TEM images for [170-700]. Fig. 2*a* shows a cross section bright field TEM image where the film thickness homogeneity can be observed. Figure 2*b* shows a dark field image obtained from the $(200)_{δ-MoN}$ ring (see selected area electron diffraction (SAED) in the inset). The bright regions in the image correspond to nanometric δ-MoN grains with typical size between 5 nm and 10 nm. In addition, the 201 reflection -corresponding to a superficial $MoO_3$ layer- is observed in the SAED [17]. Figure 2*c* shows a high resolution TEM image of δ-MoN grains. The spots corresponding to the 200 reflections are identified for one of the grains. In addition, planes at distances of ≈ 0.248 nm can be identified from Fourier filtered image. Figure 3*a* shows the AFM image for [170-700] (smother surfaces are observed for thinner films). The film displays a very smooth surface, only a few small morphologic inhomogeneities with a height less than 6 nm are observed (see profile in Fig. 3*b*). The Root Mean Square (RMS) roughness for an area of 5 µm$^2$ is 0.6 nm, which indicates that (compared to as-grown films) crystallization does not significantly alter the morphology of the films [17]. The weak influence of the annealing in the surface topology can be attributed to the crystallization of grains with diameter of a few nanometers.

XPS measurements were performed to obtain information about the chemical composition of the film and the chemical state of the Mo. The photoelectron peaks Mo3d, O1s, C1s and N1s were measured in detail. An overlap of the N1s and the Mo3p peaks was observed. In order to remove components related to superficial C and $MoO_3$, the surface of the films was cleaned with $Ar^+$ sputtering (2kV) [17]. The Mo3d binding energy region for the surface of as-grown films and [170-700] thin films is shown in Figs. 4*a* and 4*b*, respectively. The thickness of the $MoO_3$ surface layer was estimated in 1.6 nm using the XPS Mo3d intensities of the $MoO_3$ and MoN phase [20]. In the as-grown Mo3d spectra two components were identified: a major component at BE= 228.9eV attributed to the δ-MoN phase, and a minor component at BE=228.5eV attributed to the γ-$Mo_2N$ phase [21,22]. In the [170-700] films only one major component was identified at BE=228.9eV, which can be assigned to the δ-MoN phase. Besides, in both spectra there is an extra component at higher binding energies that can be ascribed to different causes: unscreened peaks of the present phases as observed for $MoO_2$ [23], the presence of $Mo^{4+}$ associated to $MoO_2$ impurities or as a consequence of overlooking possible asymmetric behavior of the main components, as suggested for $MoO_2$ [24].

The dependence of $T_c$ with the thickness for annealed films was analyzed measuring the electrical resistance as function of the temperature. To understand the influence of the thermal annealing on the electrical properties, initially a 170 nm thick MoN film was annealed at different temperatures (see Fig. 5*a*). As-grown thin films display a *RRR* ≈ 0.8 (indicative of high disorder with very short electronic mean free path *l*). After crystallization to δ-MoN (at ≈ 600 °C), the films show a metallic behavior. For [170-600] and [170-700], the *RRR* values are ≈ 1.3 and ≈ 1.5, and they display a superconducting transition with $T_c$ = 10.5 K and 11.2 K, respectively (see inset Fig. 5*a*). The superconducting transition width (defined as in Fig. 5*b*) is ≈ 1 K for [170-600] and ≈ 0.6 K [170-700]. A common feature for films with different thickness is that for longer annealing periods (> 30min) or for higher annealing temperatures (> 700°C), neither $T_c$ is increased nor the superconducting transition width is reduced. In particular, annealing temperatures higher than 800 °C produce cracks on the surface, which is evidenced in wider superconducting transitions without percolation. Figure 5*b* shows the superconducting transition observed for films with different thickness after annealing at 700 °C. The results show that $T_c$ is gradually suppressed by thickness even for films as thicker as 100 nm ($T_c$ = 10.7 K). For epitaxial δ-MoN thin films, the reduction in $T_c$ is usually observed for films thinner than 40 nm [27]. In our case, $T_c$ for $d$ = 40 nm decreases to 7.2 K, indicating that the thickness affects the crystallization. The results obtained for [170-600] and [170-700] suggest that the thickness dependence of $T_c$ may be related to changes in the disorder and in the average grain size. Detailed studies of the microstructure are necessary to clarify this point.

Following we analyzed the temperature dependence of the upper critical field $H_{c2}$ (*T*) for [170-600] and [170-700], which allows to analyze the influence of $T_c$ on $H_{c2}$ (*T*) for the 3 dimensional (3D) limit (*d* >> coherence length ξ). Figure 6 shows $H_{c2}$ (*T*) for the [170-600] and [170-700] with the magnetic field perpendicular ($H_{c2}^{\perp}$) and parallel ($H_{c2}^{//}$) to the surface. Inset in Fig. 6 shows comparison between $H_{c2}$ and the irreversibility line ($H_{irr}$, related to the superconducting transition width Δ*T*). The results show negligible upper critical field anisotropy (γ = $H_{c2}^{\perp}/H_{c2}^{//}$ ≈ 1), which is in agreement with the expectations for 3D polycrystalline thin films. The temperature dependence of $H_{c2}^{\perp}$ can be analyzed by the WHH model developed for dirty one-band superconductors [25], which is given by:

$$ln\frac{1}{t} = \sum_{v=-\infty}^{\infty}\left(\frac{1}{|2v+1|} - \left[|2v+1| + \frac{\hbar}{t} + \frac{(\alpha\hbar/t)^2}{|2v+1|+(\hbar+\lambda_{so})/t}\right]^{-1}\right) \text{ [eq. 1]},$$

where $t = T / T_c$, $\hbar = (4/\pi^2)\left(H_{c2}(T)/|dH_{c2}/dT|_{T_c}\right)$, α is the Maki parameter which quantifies the weakening influence of the Pauli electron spin paramagnetism on the superconducting state, and $\lambda_{so}$ is the spin-orbit scattering constant. The WHH formula- satisfies the relation $H_{c2}(0) = \frac{H_{c2}^{orb}(0)}{\sqrt{1+\alpha^2}}$ [eq. 2] when $\lambda_{so}$ = 0 [26]. The $H_{c2}$ (*T*) curves can be adjusted considering $\alpha = 0$, $\lambda_{so} = 0$ and $-|dH_{c2}/dT|_{T_c} \approx 1$ T/K (see dotted lines in Fig. 6). The parameter $\alpha = 0$ corresponds to a pure "orbital field limit" due to supercurrents circulating around the vortex cores. The $H_{c2}(0)$ obtained from the extrapolation to zero field of the WHH model are 7.5 T ($\xi(0) = 6.6$ nm) and 7.8 T ($\xi(0) = 6.5$ nm) for [170-600] and [170-700], respectively. These values are lower than the $\approx 10$ T previously obtained in epitaxial and dirty δ-MoN sintered for polymer assisted deposition [27]. From very basic considerations $\xi_0 \propto v_F/T_c$, where $v_F$ is the Fermi velocity [28]. Similar $\xi_0 T_c$ values ($\approx 70$ nm*K) are obtained for [170-600] and [170-700] and the films reported in Ref [27] ($\xi_0 T_c = 5.6$ nm*12.6 K), which indicates that the reduction in $H_{c2}(0)$ can be mainly attributed to the suppression of $T_c$ (corresponding to films in the dirty limit with very short electronic mean free path $l$).

Finally, for comparison with δ-MoN obtained by chemical methods [12,15], the $J_c(H)$ dependences at 4.5 K for [100-700] and [170-700] were obtained (see Fig. 7). At low fields both samples display thermal vortex jumps, which affect the magnetization (see inset Fig. 7). The $J_c(H)$ dependences are plotted for $\mu_0 H > 0.05$ T. The obtained values $J_c$ at low fields are $\approx 1.24$ MAcm$^{-2}$ [100-700] and $\approx 1.8$ MAcm$^{-2}$ [170-700], which are comparable to those reported for epitaxial thin films in refs. [12] and [15] (with values above 1 MA cm$^{-2}$ for $T = 4.5$ K].

## 4. Conclusions

In summary, we report on the synthesis and the superconducting properties in nanocrystalline δ-MoN thin films. Initially, amorphous MoN thin films were grown by reactive sputtering at room temperature on Si wafers. In a second step, the films were crystallized in vacuum to avoid surface contamination, at temperatures between 600 °C and 700 °C. The annealed films display very smooth surfaces (relevant for tunnel junctions) and a polycrystalline microstructure with nanometric grain size. The superconducting critical temperature $T_c$ is strongly affected by the film thickness. Films thicker than 100 nm display $T_c$ > 10 K, being 11.2 K for $d$ = 170 nm. The reported simple process to synthesize δ-MoN thin films with adequate physical properties on Si substrates enhances its potential for developing different technological applications, varying from radiation sensors to Josephson tunnel junctions.

**Acknowledgments**

We thank C. Olivares for technical assistance. This work was partially supported by the ANPCYT (PICT 2015-2171), U. N. de Cuyo 06/C505 and CONICET PIP 2015-0100575CO. NH, SB, MS and JG are members of the Instituto de Nanociencia y Nanotecnología, CNEA-CONICET.

Figure 1. X-ray diffraction pattern for a 170 nm MoN thin film: as grown and annealed (600 °C and 700 °C) . (*) Silicon reflections.

Figure 2. TEM images for a 170 nm thick δ-MoN film ([700]). *a)* Bright field cross section TEM image. *b)* Dark field plan view image using the $(002)_{\delta\text{-MoN}}$. Inset shows the corresponding SAED pattern. *c)* High resolution TEM image of nanograins. The fast Fourier transformation and the filtered image obtained using the 200 spots for one of the grains are included.

Figure 3. *a)* 5 x 5 μm$^2$ topographical AFM of a 170 nm thick MoN film after being annealed at 700 °C. *b)* AFM profile for the line indicated in *a)*.

Figure 4. XPS Mo3d spectra of: *a)* As-grown MoN thin films. *b)* MoN film after being annealed at 700 °C ([170-700]).

Figure 5. *a)* Temperature dependence of the resistance for an as-grown thin film, [170-400], [170-600] and [170-700]. Inset shows the temperature dependence of [170-600] and [170-700] at temperatures smaller than 20 K and the criteria for the determination of $T_c$. *b)* temperature dependence of the normalized resistance for [40-700], [60-700], [100-700] and [170-700].

Figure 6. Temperature dependence of the upper critical field ($H_{c2}$) and irreversibility line ($H_{iee}$) with **H** parallel and perpendicular to the surface for [170-600] and [170-700]. Inset shows the temperature dependence of the resistance for [170-600] and [170-700] applying different magnetic field **H** perpendicular to the surface.

Figure 7. Log-log plots of critical current density ($J_c$) versus applied magnetic field ($H$) of [100-700] and [170-700] at 4.5 K. Inset shows isothermal **M** versus **H** curves at 4.5 K applying different magnetic field **H** perpendicular to the surface.

Figure 1

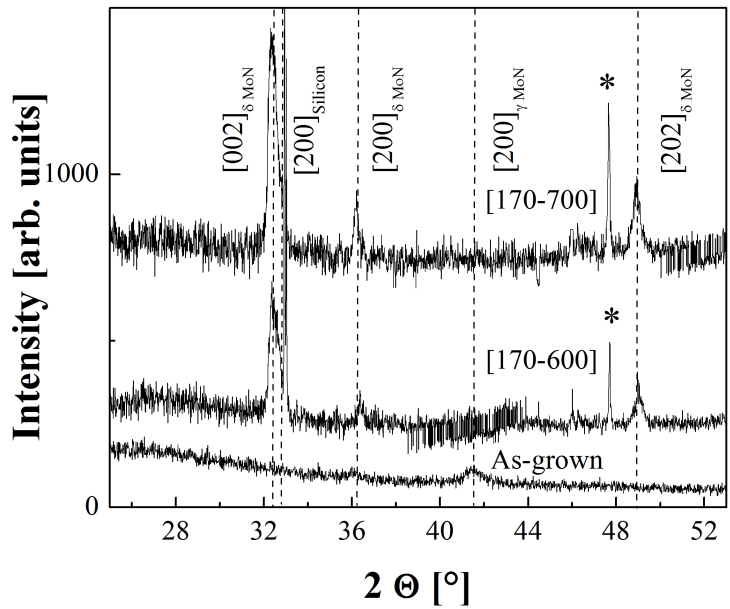

Figure 2

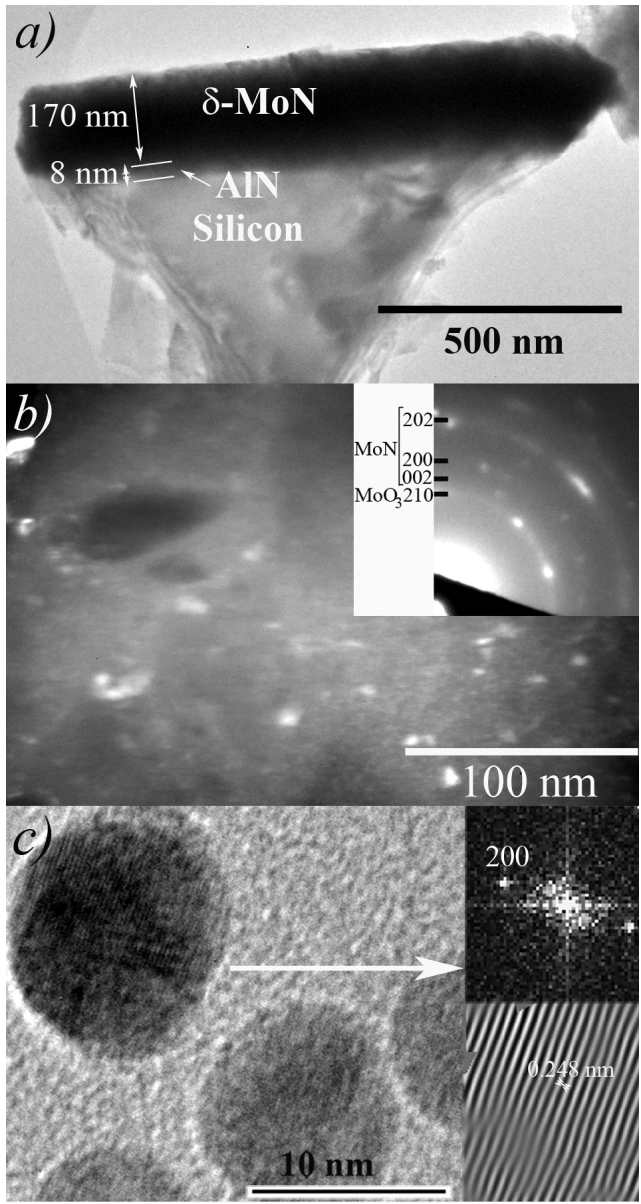

Figure 3

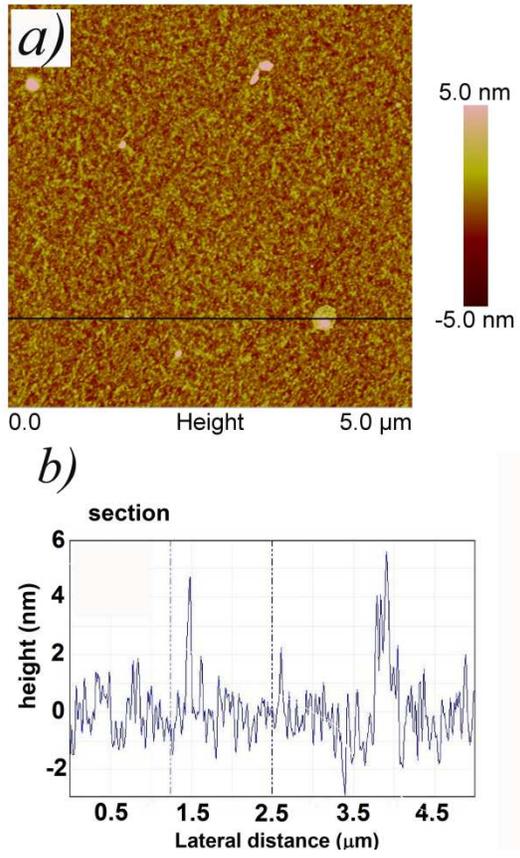

Figure 4

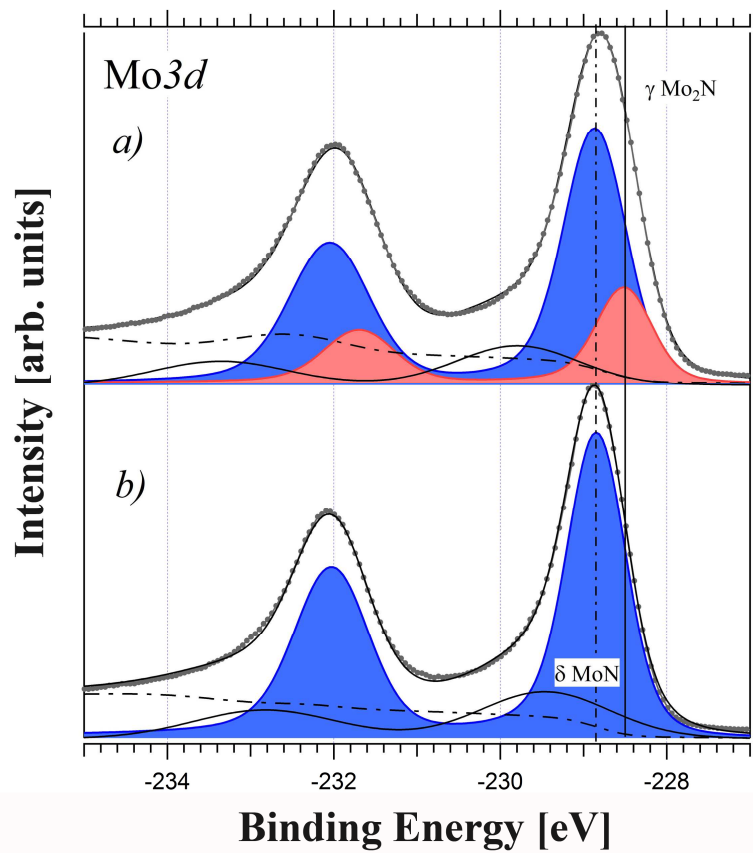

Figure 5

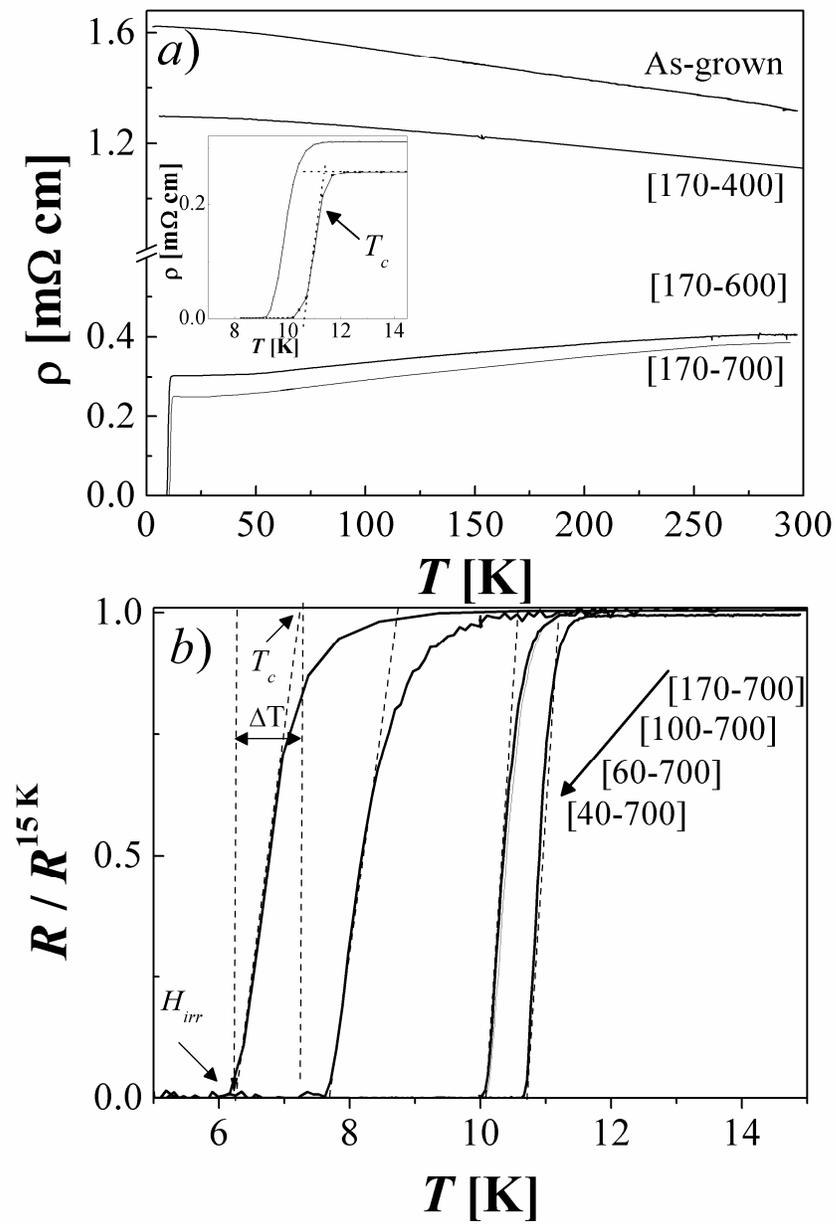

Figure 6

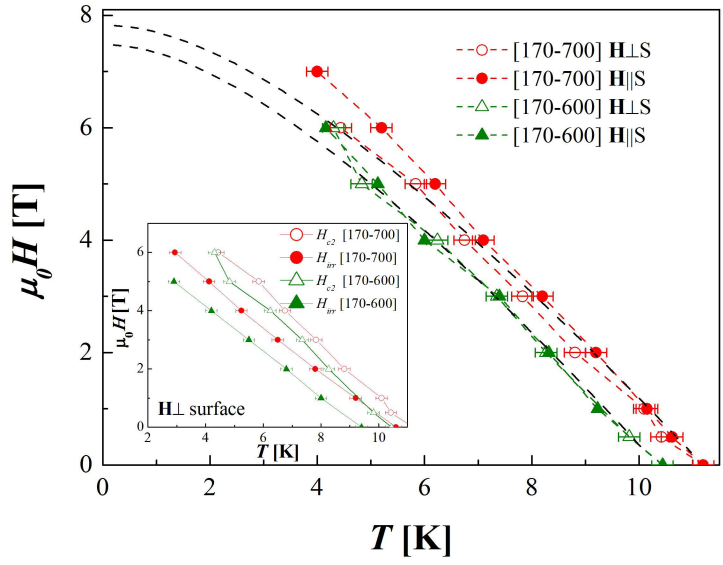

Figure 7

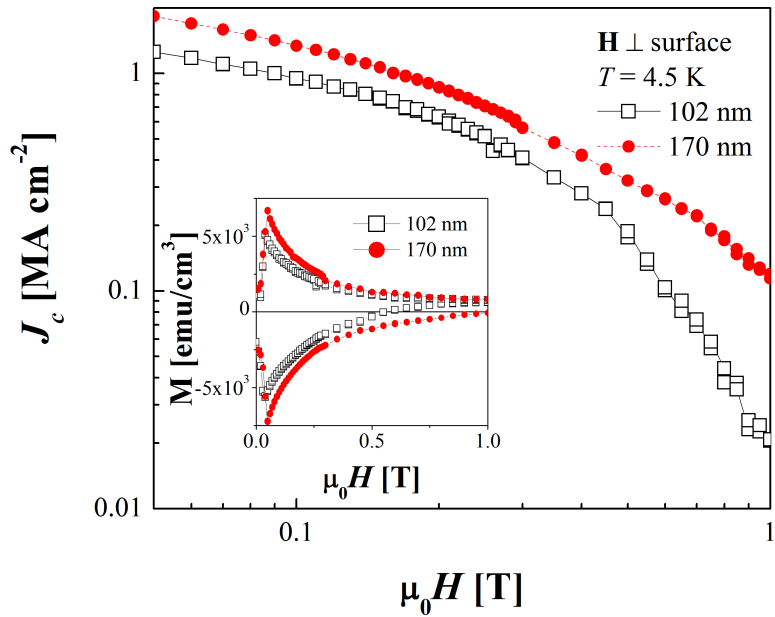